\documentstyle[aps,epsfig,preprint,tighten,float]{revtex}  
\newcommand{\be}{\begin{equation}}
\newcommand{\ee}{\end{equation}}

\begin{document}
\title{Fermionic Mapping For Eigenvalue Correlation Functions Of
Weakly Non-Hermitian Symplectic Ensemble}
\author{M. B. Hastings}
\address{Physics Department, Jadwin Hall\\
Princeton, NJ 08544\\
hastings@feynman.princeton.edu}
\maketitle
\begin{abstract}
The eigenvalues of an arbitrary quaternionic matrix have a joint probability
distribution function first derived by Ginibre.  We derive
the j.p.d. for the weakly non-Hermitan version of this problem and then
show that there exists a mapping of this system
onto a fermionic field theory.  This mapping is used
to integrate over the
positions of the eigenvalues and obtain eigenvalue density as well
as all higher correlation functions for both the strongly and
weakly non-Hermitian cases.
\end{abstract}
\section{Introduction}
Several ensembles of non-Hermitian matrices were given by Ginibre\cite{ginibre}.
These are the ensembles of matrices with arbitrary real, complex,
or quaternionic entries.  Ginibre gave joint probability distributions
for the eigenvalues for the complex and quaternionic cases, and succeeded
in obtaining correlation functions in the complex case, while correlation
functions for the quaternionic case were found later\cite{mehta}.  The purpose 
of this paper is to extend the correlation functions for the quaternionic
problem to the weakly non-Hermitian case\cite{weakly}, as well as to
introduce a fermionic mapping to simplify the computation of these correlation
functions.  Further, the mapping also permits us to derive
the 4-point, and higher, correlation functions, that were only conjectured
before.

Although Ginibre's ensembles are interesting in themselves, they are also
closely connected with the chiral random matrix ensembles that appear in
QCD\cite{qcd} and some condensed matter systems\cite{chiral}.
Knowing the eigenvalue correlation functions in the non-Hermitian ensemble,
one can easily determine correlation functions in the corresponding chiral
ensemble.  Further, the weakly non-Hermitian\cite{weakly} versions of
these ensembles are of interest in open quantum systems; there exists
a study using supersymmetric techniques of the eigenvalue
distribution in the weakly non-Hermitian
version of the ensemble considered in this paper\cite{susy}.

One interesting feature to observe in
the two-level correlation function is the crossover from a non-monotonic
correlation function with algebraic tails in the limit of very weak
non-Hermiticity to a monotonically decaying correlation function with
Gaussian tails in the limit of strong non-Hermiticity.

Another interesting property that the symplectic non-Hermitian ensemble
exhibits is a depletion of the eigenvalue density near the real axis; this
could be guessed at by looking at the joint probability distribution (j.p.d.)
derived originally by Ginibre (given in equation (\ref{jpd}) below).  The
depletion was also found numerically\cite{numerics}.

Consider an arbitrary $N$-by-$N$ matrix of quaternions.
This is equivalent to a $2N$-by-$2N$ matrix $M$ with complex entries.
Let $M$ be chosen from an ensemble of such matrices with Gaussian weight
\be
P(M)=e^{-\frac{1}{2} {\rm Tr}(M^{\dagger}M)} \, dM
\ee
This defines the strongly non-Hermitian ensemble.
The eigenvalues of $M$ come in complex conjugate pairs; for every eigenvalue
$z=x+iy$, there is an eigenvalue $\overline z=x-iy$.
Let the matrix $M$ have eigenvalues $z_i,\overline z_i$, $i=1...N$.  Then 
the j.p.d. of the eigenvalues is given, up to
a constant factor, by
\be
\label{jpd}
\frac{1}{N!}\prod \limits_{i} e^{-\overline z_i z_i} |z_i-\overline z_i|^2
\prod \limits_{i<j} 
(z_i-z_j) (z_i-\overline z_j)
(\overline z_i-z_j) (\overline z_i-\overline z_j)
\prod \limits_{i} 
{\rm d}\overline z_i\,
{\rm d}z_i
\ee
where ${\rm d}\overline z {\rm d}z=2 {\rm d}x {\rm d}y$.

In section II, a fermionic mapping is introduced to write equation
(\ref{jpd}) as a correlation function in a fermionic field theory.  The
mapping is then used to calculate the eigenvalue density.  In section
III, we introduce the weakly non-Hermitian ensemble and calculate eigenvalue
density for that ensemble.
In section IV,
multi-eigenvalue correlation functions are calculated for both strongly
and weakly non-Hermitian cases.  The calculations in section III and IV are
simple extensions of the calculation given in section II.  For this
reason, the calculation in section II is given in the most detail, while
the other calculations are sketched.

For the strongly non-Hermitian case, the main results are 
equations (\ref{singp},\ref{intrep}) for the
eigenvalue density,
and equation (\ref{2corr}) for the two-level correlation function.
For the weakly non-Hermitian case, the main results are
equation (\ref{weakden}) for the eigenvalue density and
equations (\ref{crosscor},\ref{intover}) for the two-level correlation function.
\section{Fermionic Mapping and Green's Function}
In this section we develop the fermionic mapping for the j.p.d. of
the strongly non-Hermitian ensemble.
First, we write equation (\ref{jpd}) as a correlation function
in a fermionic field theory.
Then, for convenience we shift to radial coordinates, making a conformal
transformation.  Finally, we integrate over all but one of the $z_i$ to 
obtain the eigenvalue density.  
The integral over the $z_i$ is done inside the correlation function; only
after doing the integral is the correlation function evaluated.  This amounts
to commuting the order of doing the integral and evaluating the correlation
function, and is the essential trick used in this section.
In section IV we will
demonstrate how to obtain multi-level correlation functions by a simple
extension of the procedure of this section.

\subsection{Fermionic Mapping}
First, let us show that equation (\ref{jpd}) can be written as a correlation 
function
in a two-dimensional fermionic field theory.  A similar fermionic
mapping was demonstrated previously for the Hermitian orthogonal and
symplectic ensembles\cite{me}.  Let the field $\psi(x)$ have
have the action
\be
S=\frac{1}{2}\int {\rm d}\overline z \, {\rm d}z \psi^{\dagger}(z) \overline 
\partial \psi(z)
\ee
Note that we are using only one chirality of fermionic field.
Consider a correlation function of this field, such as
\be
\langle \prod \limits_{i=1}^{2N} \psi(a_i) \psi^{\dagger}(b_i) \rangle
\ee
This correlation function is equal to
\be
\label{crfn}
\frac{ \Bigl( \prod \limits_{i<j}^{2N} (a_j-a_i) \Bigr)
\Bigl( \prod \limits_{i<j}^{2N} (b_j-b_i) \Bigr)}
{\Bigl( \prod \limits_{i,j}^{2N} (a_j-b_i) \Bigr)}
\ee
Let us consider a specific choice of $b_j$, with 
$b_j=L e^{2\pi i \frac{j}{2N}}$.  In the limit $L \rightarrow \infty$, we
find that equation (\ref{crfn}) reduces to
\be
\label{crsimp}
L^{-2N^2-N} \prod \limits_{i<j}^{2N} (a_j-a_i)
\ee
Comparing this to equation (\ref{jpd}) we realize that equation (\ref{jpd})
can be written as
\be
\label{cfin}
\frac{1}{N!}
\lim\limits_{L \rightarrow\infty} 
L^{2N^2+N}
\langle
\Bigl(
\prod\limits_{j=1}^{2N}
\psi^{\dagger}(b_j)
\Bigr)\,
\Bigl(
\prod \limits_{j=i}^{N} 
U(\overline z_j,z_j)
\psi(z_j) \psi(\overline z_j)
{\rm d}\overline z_j \, {\rm d}z_j 
\Bigr)
\rangle
\ee
where $b_j=L e^{2\pi i \frac{j}{2N}}$.
and $U(\overline z,z)=e^{-\overline z z} (z-\overline z)$.
When integrating over $z_i$, the limit $L\rightarrow\infty$ must be
taken before doing the integral over $z_i$.

Now we will make a conformal transformation to radial coordinates.  Write
$z=e^w$ and $\overline z=e^{\overline w}$.  Let $w=t+i\theta$.  
The action for the
fermionic field is unchanged under this transformation, but we must change
equation (\ref{cfin}) as the field $\psi$ has non-vanishing scaling dimension
and conformal spin.  Equation (\ref{cfin}) gets replaced by
\be
\label{radeq}
\frac{1}{N!}
\lim\limits_{L \rightarrow\infty} 
L^{2N^2}
\langle
\Bigl(
\prod \limits_{j=1}^{2N}
\psi^{\dagger}(v_j)
\Bigr)\,
\Bigl(
\prod \limits_{j=1}^{N}
e^{t_j} 
U(e^{\overline w_j},e^{w_j}) 
\psi(w_j) \psi(\overline w_j)
{\rm d}\overline w_j \, {\rm d}w_j 
\Bigr)
\rangle\ee
where $v_j=\log{b_iL} +2\pi i \frac{j}{2N}$.

Now, we will introduce Fourier transforms for the creation and annihilation
operators.  We will write $\psi(w_i)=\sum_{k}e^{k w_i}a(k)$ and
$\psi^{\dagger}=\sum_{k}e^{-k w_i}a^{\dagger}(k)$.  
In the limit 
$L\rightarrow\infty$, the only states involved in equation (\ref{radeq})
are those with $k=1/2,3/2,5/2,...,2N-1/2$.  If there are excitations
in states with higher $k$, they will vanish in the large $L$ limit.

Then, we can rewrite equation (\ref{radeq}), up to factors of order unity, as
\be
\label{anieq}
\frac{1}{N!}
\langle
\Bigl(
\prod\limits_{k=1/2}^{2N-1/2}a^{\dagger}(k)\Bigr)\,
\prod\limits_{j=1}^{N}\Bigl(\sum\limits_{k,k'}
a(k)a(k')
e^{w_j k}e^{\overline w_j k'}
\,
e^{t_j} 
U(e^{\overline w_j},e^{w_j}) 
{\rm d}\overline w_j \, {\rm d} w_j
\Bigr)
\rangle
\ee

In equation (\ref{anieq}), consider integrating over $w_j,\overline w_j$ for
some given set of $j=1...M$.  We will do the integral inside the
correlation function.
The integral
\be
\int
\prod\limits_{j=1}^{M}\Bigl(\sum\limits_{k,k'}
a(k)a(k')
\,
e^{w_j k}e^{\overline w_j k'}
e^{t_j} 
U(e^{\overline w_j},e^{w_j}) 
{\rm d}\overline w_j \, {\rm d} w_j\Bigr)
\ee
is equal to
\be
\label{fouran}
O^{M}
\ee
where the operator $O$ is defined by
\be
O=\sum_{k}a(k)a(k+1) 4\pi(k+1/2)!
\ee
Therefore, if we integrate over all eigenvalues in equation (\ref{anieq}),
we obtain
\be
\label{normalize}
Z=\frac{1}{N!}\langle \Bigl( \prod\limits_{k=1/2}^{2N-1/2}a^{\dagger}(k)\Bigr)\,
O^{N} \rangle =
\prod_{k} \Bigl( 4\pi (k+1/2)!\Bigr)
\ee
where the product extends over $k=1/2,5/2,9/2,...,2N-3/2$.

\subsection{Eigenvalue Density}
To calculate the density of eigenvalues, we must integrate over all, except
one, of the coordinate pairs $\overline w_i,w_i$ in equation (\ref{anieq}).
Using equation (\ref{fouran}), and normalizing with equation
(\ref{normalize}), we wish to compute
\be
\label{onem}
\frac{
U(e^{\overline w},e^{w})
e^t
\sum\limits_{m,m'}
e^{m w}e^{m' \overline w}
\,
\langle
\Bigl(\prod\limits_{k=1/2}^{2N-1/2}a^{\dagger}(k)\Bigr)
\,
O^{N-1}
\,
a(m)a(m')
\rangle
{\rm d}\overline w \, {\rm d} w}
{(N-1)!Z}
\ee
It may be verified that the correlation function
appearing in the sum of equation 
(\ref{onem}) is nonvanishing only if either
$m-1/2$ is even, $m'-1/2$ is odd, and $m<m'$, or if $m'-1/2$ is even,
$m-1/2$ is odd, and $m'<m$.  In the first case, with $m<m'$, the contribution
to equation (\ref{onem}) is
\be
\frac{1}{4\pi} \prod_{k}\frac{1}{(k+1/2)!}
\prod_{l}(l+1/2)!
e^{-\overline z z} \sqrt{\overline z z}(z-\overline z)
z^{m}\overline z^{m'}
{\rm d}\overline w \, {\rm d} w
\ee
where the product over $k$ extends over 
\be
k=1/2,5/2,9/2,...
\ee
and the product over $l$ extends over 
\be
l=1/2,5/2,9/2...m-2,m+1,m+3,...,m'-2,m'+1,m'+3,...,2N-1/2
\ee
This is equal to
\be
\label{firstc}
\frac{1}{4\pi}
\frac{1}{(m-1/2)!!(m'-1/2)!!}
e^{-\overline z z} \sqrt{\overline z z}(z-\overline z)
z^{m}\overline z^{m'}
{\rm d}\overline w \, {\rm d} w
\ee

In the second case, with $m'<m$, the result is
\be
\label{secondc}
-\frac{1}{4\pi}\frac{1}{(m-1/2)!!(m'-1/2)!!}
e^{-\overline z z} \sqrt{\overline z z}(z-\overline z)
z^{m}\overline z^{m'}
{\rm d}\overline w \, {\rm d} w
\ee

We can obtain the eigenvalue density $\rho(\overline z,z)$ by
adding equations (\ref{firstc},\ref{secondc}) and summing over $m,m'$.
Shifting $m$ and $m'$ by one-half, and changing from 
${\rm d}\overline w \, {\rm d} w$ to
${\rm d}\overline z \, {\rm z} w$,
we find that the final result for the eigenvalue density $\rho(\overline z,z)
{\rm d}\overline z \, {\rm d}z$
is
\be
\label{singp}
\rho(\overline z,z) {\rm d}\overline z \, {\rm d}z
=\frac{1}{4\pi}e^{-\overline z z}(z-\overline z)
G(\overline z,z){\rm d}\overline z \, {\rm d}z
\ee
where the Green's function $G(\overline z,z)$ is given by
\be
\label{gnfn}
G(\overline z,z)=
\sum\limits_{m<m';m=0,2,4,...;m'=1,3,5,...}
\frac{1}{m!!m'!!}
(z^m \overline z^{m'}- \overline z^m z^{m'})
\ee

\subsection{Discussion}
Let us now look at the properties of equation (\ref{singp}).  We will
discuss in turn the normalization of the density; the way the density
depends on $x$ and $y$ separately, where $z=x+iy$; an integral
representation for the density; the circular law;
and the depletion of density near the real axis.

First, consider the normalization of the single particle density.
It is automatic from the above derivation that the eigenvalue density
is properly normalized, although one must be careful about defining
the normalization depending on whether one is counting total
number of eigenvalues or total number of pairs of eigenvalues.
The normalization is defined such that
$\int \rho(\overline z,z) {\rm d}\overline z {\rm d}z=N$.

Next, writing $z=x+iy$ and $\overline z=x-iy$,
one can show by differentiating the power series in equation (\ref{gnfn}) that 
$\partial_{x}G(x+iy,x-iy)=
\frac{\partial+\overline \partial}{2}G(x+iy,x-iy)=
2x G(x+iy,x-iy)$, for large $N$.
This implies that 
\be
\label{faceq}
G(x+iy,x-iy)=e^{x^2} f(y)
\ee
and therefore
$\rho=\frac{1}{4\pi}2yi e^{-y^2} f(y)$, for some function $f$, so the
interesting properties of the eigenvalue density are contained in $f(y)$.
Later we will discuss the properties of $f(y)$ for small $y$, and show
that there is a depletion of the density of eigenvalues near the
real axis.

Using equations (\ref{singp},\ref{gnfn},\ref{faceq}), 
we can derive an integral representation
for $\rho$.  We can use equation (\ref{faceq}) to write
\be
G(\overline z,z)=e^{\overline z z}G(\overline z-z,0)
\ee
Then equation (\ref{gnfn}) implies that
\be
G(\overline z-z,0)=\sum_{m=1,3,5,...} \frac{1}{m!!} (\overline z-z)^m
\ee
This is equal to
\be
\int\limits_{0}^{\infty} i\, {\rm Sin}(\frac{\overline z-z}{i}t) e^{-t^2/2} 
\, {\rm d}t
\ee
Using this integral representation in equation (\ref{singp}) we find that
\be
\label{intrep}
\rho(\overline z,z) 
=\frac{1}{4\pi}2y
\int\limits_{0}^{\infty} {\rm Sin}(2yt) e^{-t^2/2} \, 
{\rm d}t
\ee

Let us now consider the circular law.  For large $y$, equation
(\ref{intrep}) reduces to
\be
\rho(\overline z,z) {\rm d}\overline z \, {\rm d}z
\rightarrow \frac{1}{4\pi} {\rm d}\overline z \, {\rm d}z
\ee
So, the density tends to a constant for large $y$.
However, this integral representation is valid only for $N$ infinite;
for finite $N$,
the density tends to a constant only within a disc of radius $\sqrt{2N}$, and 
vanishes outside the
disc.  This is the well-known circular law\cite{ginibre,circle}.  
The vanishing of the density outside the disc is 
easy to see from the power series representation.  For
finite $N$ the highest power of $(\overline z z)$ appearing
in equation (\ref{gnfn}) is roughly $2N$ and so $\rho$ will be
exponentially small for $(\overline z z)>2N$.  

Note that the total density in the disc is correct.  The area of a
disc of radius R is $2 \pi R^2$, where we are using the measure
${\rm d}\overline z \, {\rm d}z=2 {\rm d}x \, {\rm d}y$.  
The density is $\frac{1}{4 \pi}$.  So,
the number of particles in a disc of radius $\sqrt{2N}$ is 
indeed $N$, as desired.

For small $y$, we find that $\rho$ is reduced below the expected
result.  Such a reduction was found numerically before\cite{numerics}.
In the figure, we graph the eigenvalue density as a function of $y$, for
$x=0$, for a system of 100 particles.
\section{Weakly Non-Hermitian Case}
Now we will consider the weakly non-Hermitian version of the ensemble
given above.  In the weakly non-Hermitian random matrix ensemble,
we again consider an arbitrary $N$-by-$N$ matrix of quaternions, $H$, but
use a different Gaussian weight.  Let
$H=H_h+H_a$, where the $H_h$ is Hermitian and $H_a$ is anti-Hermitian. 
Then, we chose the matrix $H$ with Gaussian weight
\be
\label{chosen}
e^{-\frac{N}{2}{\rm Tr}(H_h^{\dagger}H_h)-\frac{N^2 a}{2} {\rm Tr}(H_a^{\dagger}
H_a)}
\ee
where $a$ is some constant.  In the large $a$ limit, this
reduces to the Gaussian Symplectic Ensemble.  For finite $a$, 
the weight in equation (\ref{chosen}) is chosen to make sure that
the imaginary part of the eigenvalues scales with $N$ in the
same way as the level spacing.

If the matrix $H$ is chosen with weight given by equation (\ref{chosen}),
then the j.p.d. of equation (\ref{jpd}) gets replaced by
\be
\label{jpdw}
\frac{1}{N!}\prod \limits_{i} e^{-N x_i^2-N^2 a y_i^2} 
|z_i-\overline z_i|^2
\prod \limits_{i<j} 
(z_i-z_j) (z_i-\overline z_j)
(\overline z_i-z_j) (\overline z_i-\overline z_j)
\prod \limits_{i} 
{\rm d}\overline z_i\,
{\rm d}z_i
\ee
The only difference in the weakly non-Hermitian case
is that the Gaussian function of eigenvalue position
$e^{-\overline z_i z_i}$ is replaced by $e^{-x_i^2-Na y_i^2}$.

We have not found equation (\ref{jpdw}) previously in the literature.
This equation can be derived most easily as follows: write an $N$-by-$N$
matrix of quaternions, $H$, as
\be
H=X^{-1} T X
\ee
where $X$ is a quaternion matrix such that $X^{-1}=X^{\dagger}$,
and $T$ is an upper triangular matrix of quaternions.  
This procedure is a Schur decomposition, and
is possible since the field of quaternions, like the
field of complex numbers, is algebraically closed.

The eigenvalues, $z_i$, can be obtained
from the diagonal elements of $T$; each diagonal element of $T$ is
a quaternion, which is associated with a pair of complex conjugate eigenvalues
$z_i,\overline z_i$.
If a given diagonal element of $T$ is $T_i=A+Bi+Cj+Dk$, then
$z_i=A\pm i \sqrt{B^2+C^2+D^2}$.

The Jacobian associated with
this change of variables is $\prod \limits_{i < j} 
(z_i-z_j) (z_i-\overline z_j)$.  
Further,
\be
e^{-\frac{N}{2}{\rm Tr}(H_h^{\dagger}H_h)-\frac{N^2 a}{2} {\rm Tr}(H_a^{\dagger}
H_a)}=
e^{-\frac{N}{2}{\rm Tr}(T_h^{\dagger}T_h)-\frac{N^2 a}{2} {\rm Tr}(T_a^{\dagger}
T_a)}
\ee
where $T_h,T_a$ are Hermitian and anti-Hermitian parts of $T$.
The integral over the elements of
$T$ above the diagonal can be done trivially as this integral
is Gaussian.  The integral over the
diagonal elements of $T$ includes a Gaussian factor and a factor from
the Jacobian.  This integral is exactly the integral over the j.p.d. of
equation (\ref{jpdw}).  

Given equation (\ref{jpdw}), we could follow the procedure of the previous
section.  However, we would run into some difficulties which are purely
technical.  The problem is that, while in the strongly
non-Hermitian case the eigenvalue density is independent of $x$,
it is not independent of $x$ in the weakly non-Hermitian case.
This makes the power series expansion very awkward.  We will find
it convenient to change to a different geometry, given in
equation (\ref{dc}) below, for the weakly non-Hermitian
case.  Let me again stress that the reason for choosing a different
geometry is purely technical, to simplify the math.

An analogous simplification is often used in the Hermitian ensembles.  
For example,
consider the Gaussian Symplectic Ensemble in the large $N$ limit.  The
eigenvalue density is a function of energy, but if one appropriately
scales all energies by the local level spacing, it is simpler to obtain
correlation functions from the Circular Symplectic Ensemble\cite{dyson}.

Let us introduce new coordinates, $z=\phi+ir$ and $\overline z=\phi-ir$, 
where $\phi$ is periodic
with period $2\pi$.  Let us replace equation (\ref{jpdw}) by
\be
\label{dc}
\frac{1}{N!}\prod \limits_{i} e^{-a N^2 r_i^2} 
\frac{(e^{i z_i}-e^{i \overline z_i})^2}
{e^{-2 i \phi_i}}
\prod \limits_{i<j} 
\frac{(e^{i z_i}-e^{i z_j})
(e^{i z_i}-e^{i \overline z_j})
(e^{i \overline z_i}-e^{i z_j})
(e^{i \overline z_i}-e^{i \overline z_j})}
{e^{-2 i \phi_i-2 i \phi_j}}
\prod \limits_{i} 
{\rm d}\overline z_i\,
{\rm d}z_i
\ee
This describes a system of $N$ pairs of levels, with average level
spacing $2\pi/N$.  The imaginary part of the level is of order $1/N$, so
it is of order the level spacing.  For large $a$ this reduces to the Circular
Symplectic Ensemble.  For finite $a$ and large $N$,
we expect that the ensemble of equation (\ref{dc}) reproduces the behavior of 
the ensemble of equation (\ref{jpdw}) within a small neighborhood of some given
energy, just as the Circular Symplectic Ensemble reproduces the results
of the Gaussian Symplectic Ensemble within a neighborhood of a given energy.

The next step is to write equation (\ref{dc}) as a correlation function
in a fermionic field theory.   We will introduce $N$ creation operators
at $r=+\infty$ and $N$ creation operators at $r=-\infty$.
We find that the desired correlation function
is
\be
\label{cc1}
\frac{1}{N!}
\lim\limits_{L\rightarrow\infty} e^{N^2L} \langle
\Bigl(\prod\limits_{j=1}^{N}\psi^{\dagger}(b_j)\Bigr)\,
\Bigl(\prod\limits_{j=1}^{N}\psi^{\dagger}(c_j)\Bigr)\,
\Bigl(\prod \limits_{j} U(r_j)
\psi(\overline z_j)\psi(z_j)
{\rm d}\overline z_i\, {\rm d}z_i \Bigr)
\rangle
\ee
where $b_j=\frac{2\pi j}{N}+iL$ and $c_j=\frac{2\pi j}{N}-iL$
and $U(r_j)=e^{-a N^2 r_j^2} (e^{r_j}-e^{-r_j})$.
For large $N$, we can write $U(r_j)=e^{-a N^2 r_j^2} 2r_j$.

Now, we will introduce Fourier modes for the creation and annihilation
operators, writing $\psi(z)=\sum_{k}e^{i k z}a(k)$ and
$\psi^{\dagger}(z)=\sum_{k}e^{-i k z}a(k)$.
In the limit 
$L\rightarrow\infty$, the only states involved in equation (\ref{cc1})
are those with $k=-N+1/2,-N+3/2,...,N-1/2$.
If there are excitations
in states with higher $k$, they will vanish in the large $L$ limit.
Then, equation (\ref{cc1})
can be written as
\be
\label{cfour}
\frac{1}{N!}
\lim\limits_{L\rightarrow\infty} \langle
\Bigl(\prod\limits_{k=-N+1/2}^{N-1/2}a^{\dagger}(k)\Bigr)\,
\prod \limits_{j} 
\Bigl(\sum \limits_{k,k'}a(k)a(k')
e^{i k \overline z_j}e^{i k' z_j}
U(r_j)
{\rm d}\overline z_i\, {\rm d}z_i \Bigr)
\rangle
\ee

As in the previous section, we will integrate over some set of $z_j$, for
$j=1...M$, inside the correlation function.  The integral 
\be
\int \prod \limits_{j=1}^{M} 
\Bigl(\sum \limits_{k,k'}a(k)a(k')
e^{i \overline k z_j}e^{i k' z_j}
U(r_j)
{\rm d}\overline z_j\, {\rm d}z_j \Bigr)
\ee
is equal to
\be
\label{ow}
O_w^M
\ee
where the operator $O_w$ is defined by
\be
O_w=
8\Bigl(\frac{\pi}{aN^2}\Bigr)^{3/2} \sum \limits_{k} 
\Bigl(k e^{\frac{k^2}{aN^2}} a(k) a(-k) \Bigr)
\ee

So, if we integrate over all coordinates $\overline z,z$ in equation
(\ref{cfour}), we obtain
\be
\label{zw}
Z=\frac{1}{N!}\langle
\Bigl(\prod\limits_{k=-N+1/2}^{N-1/2}a^{\dagger}(k)\Bigr)\,
(O_w)^N \rangle=
\prod\limits_{k=1/2}^{N-1/2} \Bigl(
16(\frac{\pi}{aN^2})^{3/2}
k e^{\frac{k^2}{aN^2}}
\Bigr)
\ee
To obtain the eigenvalue density, we must integrate over all but
one of the coordinates in equation (\ref{cfour}).  Using equation (\ref{ow}),
and normalizing with equation (\ref{zw}), we obtain
\be
\label{inter}
\frac{
U(r)
\sum \limits_{m,m'}
e^{i m \overline z}e^{i m' z}
\langle
\Bigl(\prod\limits_{k=-N+1/2}^{N-1/2}a^{\dagger}(k)\Bigr)\,
(O_w)^{N-1} a(m) a(m')\rangle 
{\rm d}\overline z\, {\rm d}z}
{(N-1)!Z}
\ee
The correlation function in equation (\ref{inter})
is non-vanishing only if $m=-m'$.  Equation
(\ref{inter}) is equal to
\be
\label{nex}
U(r)
\sum\limits_{m=-N+1/2}^{N-1/2}
e^{i m (\overline z-z)} G_w(m)
{\rm d}\overline z\, {\rm d}z
\ee
where 
\be
G_w(m)=
\frac{1}{16 m} (\frac{\pi}{aN^2})^{-3/2} e^{-\frac{m^2}{aN^2}}
\ee

In the large $N$ limit, we can simplify equation (\ref{nex}) by
introducing scaled coordinates.  Let us introduce $k=m/N$ and let us also
scale $z$ by a factor of $N$ so that $\phi$ now runs from $0$ to $2\pi N$.
Then we can replace the sum by an integral and obtain
\be
\label{weakden}
\rho(\phi,r) {\rm d}\phi \, {\rm d}r=\frac{1}{4} (\frac{\pi}{a})^{-3/2}
r e^{-a r^2} G_w(\overline z,z) {\rm d}\phi \, {\rm d}r
\ee
where
\be
G_w(\overline z,z)=\int\limits_{-1}^{1} 
e^{i k (\overline z-z)} e^{-k^2/a} \frac{1}{k}
{\rm d}k
\ee

As in the previous section, the proper normalization of the above result is 
automatic from the
derivation.  It is possible to show that equation (\ref{weakden}) is 
equivalent to equation (\ref{intrep}) in the limit of very small $a$.
The qualitative feature of a depletion of eigenvalues near the real
axis is the same for weak and strong non-Hermiticity.
Equation (\ref{weakden}) may be compared to the results of the SUSY 
calculation\cite{susy}, and found to agree, with some differences in
notation between the two calculations. 
\section{Multi-Point Correlation Functions}
The calculation of multi-level correlation functions is quite easy.  In
equations (\ref{anieq},\ref{cfour}), we must integrate over all except 
for two, three,
or more, of the coordinate pairs $\overline w_i,w_i$.  Since the
system is a non-interacting fermion system, the multi-point
Green's functions can be expressed very simply in terms of the Green's
function (\ref{gnfn}), using Wick's theorem.  This permits the two-point
correlation function to be easily generalized to a multi-point correlation
function, as conjectured previously\cite{mehta}.
We will not show this in detail, but simply
sketch the results, first for the strongly non-Hermitian case and then
for the weakly non-Hermitian case.

\subsection{Strongly Non-Hermitian Case}
First let us examine the strongly non-Hermitian case, generalizing the
results of section II.
Consider the two-level correlation function, the probability to
find a pair of levels
at position $\overline z,z$ given that there is another pair
at position $\overline z',z'$.  Then, the two-level correlation function is
\be
\label{2corr}
\bigl(\frac{1}{4\pi}\bigr)^2
e^{-\overline z z-\overline z' z'} 
(z-\overline z)
(z'-\overline z')
\Bigl(
G(\overline z,z)G(\overline z',z')
-G(\overline z,z')G(\overline z',z)
+G(\overline z,\overline z')G(z',z)
\Bigr)
{\rm d}\overline z \, {\rm d}z \, {\rm d}\overline z' \, {\rm d}z'
\ee
This is just an application of Wick's theorem.

Let us consider the behavior of equation (\ref{2corr}) in the limit
when both $z$ and $z$ are far from the real axis.  Without loss of generality,
assume that ${\rm Re}(z')>{\rm Re}(z)$.  Then, use the integral
representation of the Green's function to rewrite 
$e^{-\overline z z-\overline z' z'}
G(\overline z,z')G(\overline z',z)$ as
\be
\label{theq}
e^{-|z-z'|^2}
\Bigl(
\int\limits_{0}^{\infty}e^{(\overline z-z')t}e^{-t^2/2}{\rm d}t
-\frac{1}{2}e^{(\overline z-z')^2/2}
\Bigr)
\Bigl(
\frac{1}{2}e^{(\overline z'-z)^2/2}
-\int\limits_{-\infty}^{0}e^{(\overline z'-z)t}e^{-t^2/2}{\rm d}t
\Bigr)
\ee
We can find a similar representation for $G(\overline z,\overline z') G(z',z)$.
Now, in the limit with  $z$ and $z'$ both far from the real axis, then either
${\rm Im}(z-z')$ is large or ${\rm Im}(z-\overline z')$ is large.  In the
first case, equation (\ref{theq}) is exponentially small because
of the factor of $e^{-|z-z'|^2}$.  In the second
case, the integrals over $t$ can be performed in this limit, while
$e^{(\overline z-z')^2/2}$ and
$e^{(\overline z'-z)^2/2}$ are small.  The integral,
$\int\limits_{0}^{\infty}e^{(\overline z-z')t}e^{-t^2/2}{\rm d}t$, is
equal to $\frac{1}{z'-\overline z}$, for large ${\rm Im}(z'-\overline z)$;
here we rely on the fact that ${\rm Re}(z'-\overline z)>0$.

So, up to exponentially small terms,
equation (\ref{theq}) is equal to
\be
\label{testeq}
e^{-|z-z'|^2}
\frac{1}{\overline z-z'}\frac{1}{\overline z'-z}
\ee
Also, in this limit, if equation (\ref{testeq}) is not exponentially
small, then $\frac{1}{\overline z-z'}\frac{1}{\overline z'-z}=
\frac{1}{\overline z-z}\frac{1}{\overline z'-z'}$.
Inserting this result, and similar results for 
$G(\overline z,\overline z') G(z',z)$,
back into equation (\ref{2corr}) we find that, for both
$z$ and $z'$ far from the real axis, the two-level correlation function
is equal to
\be
\label{eqa}
\bigl(\frac{1}{4\pi}\bigr)^2
\Bigl(1 -e^{-|z-z'|^2} -e^{-|z-\overline z'|^2}\Bigr)
{\rm d}\overline z \, {\rm d}z \, {\rm d}\overline z' \, {\rm d}z'
\ee
This is essentially the same as the two-level correlation function
found in the complex non-Hermitian case\cite{ginibre}.

For $z$ and $z'$ near the real axis, I have examined the behavior
of equation (\ref{2corr}) numerically.  If ${\rm Im}(z)={\rm Im}(z')$,
then the correlation function is a monotonically decaying function
of ${\rm Re}(z-z')$, with no signs of any oscillation.  The correlation
function is exponentially small if both $z$ and $z'$ are near the
real axis.
\subsection{Weakly Non-Hermitian Case}
Now let us consider the two-level correlation function in the weakly
non-Hermitian case, the probability to find one pair of levels
$\overline z,z$ given that there is another pair at
$\overline z',z'$.  As before, we must integrate over all except
for two of the eigenvalue coordinates.  Using the scaled coordinates,
we find that the two-level
correlation function is given by
\be
\label{crosscor}
\frac{1}{16} (\frac{\pi}{a})^{-3}
e^{-a r^2-a r'^2} 
rr'
\Bigl(
G_w(\overline z,z)G_w(\overline z',z')
-G_w(\overline z,z')G_w(\overline z',z)
+G_w(\overline z,\overline z')G_w(z',z')
\Bigr) 
{\rm d}\phi \, {\rm d}r
{\rm d}\phi' \, {\rm d}r'
\ee
As in the previous subsection, this is just an application of Wick's theorem.

To examine the behavior of the two-level correlation function, let us
integrate over $r,r'$ in equation (\ref{crosscor}), to be 
left with a function of $\phi-\phi'$.
The result is
\be
\label{intover}
\frac{1}{4\pi^2} -
\frac{1}{32\pi^2}
\int\limits_{-1}^{1}
\frac{(k+k')^2}{k k'}
e^{-(k-k')^2/(2a)}
e^{ik(\phi-\phi')}e^{ik'(\phi'-\phi)}
{\rm d}k {\rm d}k'
\ee
In the limit $a \rightarrow \infty$, the integral over $k,k'$ in the above 
expression can be performed to yield
\be
\label{csk}
\frac{1}{4\pi^2} -
\frac{1}{4\pi^2}
\int\limits_{-2}^{2}
\Bigl(1-\frac{|k|}{2}-\frac{|k|}{4}{\rm log}(|k|-1)
e^{ik(\phi-\phi')}\Bigr)
{\rm d}k 
\ee
Equation (\ref{csk}) is the known result for the correlation function in 
the Circular
Symplectic Ensemble.  It is a non-monotonic function, algebraically
decaying for large $\phi$.  For sufficiently small $a$, equation
(\ref{intover}) will describe a monotonically decaying function
of $\phi'-\phi$, but for fixed, non-vanishing $a$, the function
will always decay algebraically for large $\phi$.
\section{Conclusion}
In conclusion, we have given a simple fermionic mapping for determining
the correlation functions of the non-Hermitian symplectic ensemble.  
Although the eigenvalue density was found previously using SUSY, the
present derivation is simpler and can be more easily extended to the
two-level correlation function.  The two-level correlation function in 
the strongly non-Hermitian case was found to be similar to that for the 
ensemble of arbitrary complex matrices.  In the weakly non-Hermitian
case, the two-level correlation function exhibits an interesting crossover
as a function of $a$.

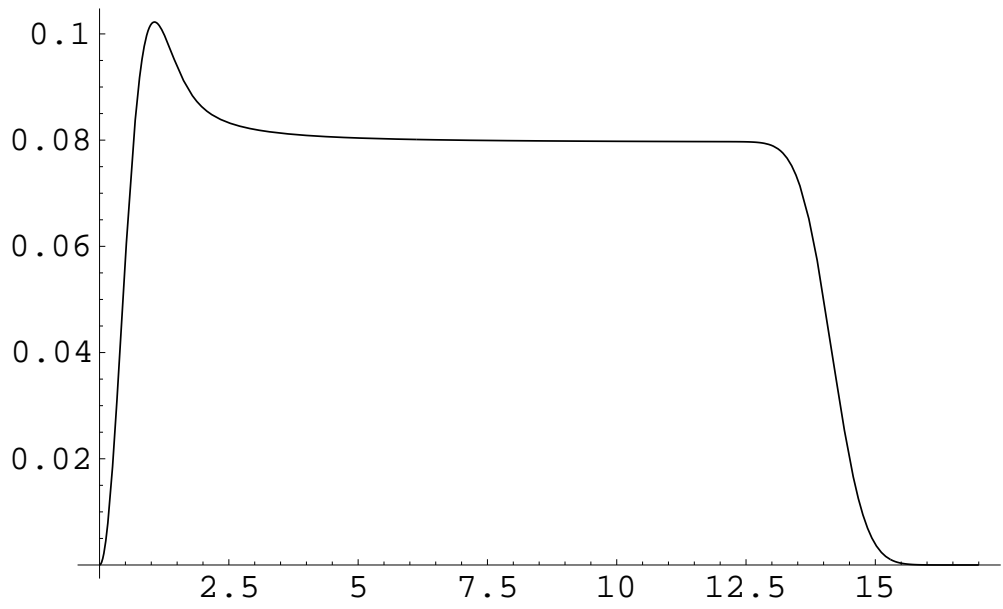
\end{document}